\documentclass[aps,pra,12pt,showkeys,superscriptaddress,reprint,longbibliography]{revtex4-1}
\usepackage{amsmath}
\usepackage{amsthm}
\usepackage{graphicx}
\usepackage[colorlinks]{hyperref}
\usepackage{dcolumn}
\allowdisplaybreaks
\usepackage[usenames,dvipsnames,table]{xcolor}
\usepackage{fancyhdr}
\usepackage{mathptmx}
\usepackage[small,raggedright]{titlesec}
\usepackage{wrapfig}
\newcolumntype{M}{D{.}{.}{1.8}}

\begin{document}

\thispagestyle{plain}

\title
{
\textcolor{BlueViolet}{
Determination of electron-hole correlation length in CdSe quantum dots
using explicitly correlated two-particle cumulant}
}

\author{Christopher J. Blanton}

\author{Arindam Chakraborty}
\email[corresponding author: ]{archakra@syr.edu}
\affiliation 
{
Department of Chemistry, Syracuse University, Syracuse, New York 13244
}

\date{\today} 

\begin{abstract} 
\noindent\textcolor{BlueViolet}{\rule{14.6cm}{1.2pt}}
        \colorbox{Apricot}{\parbox{0.8\linewidth}{%
\begin{wrapfigure}{r}{1.75in}
\vspace{-10 pt}
\includegraphics[width=1.75in]{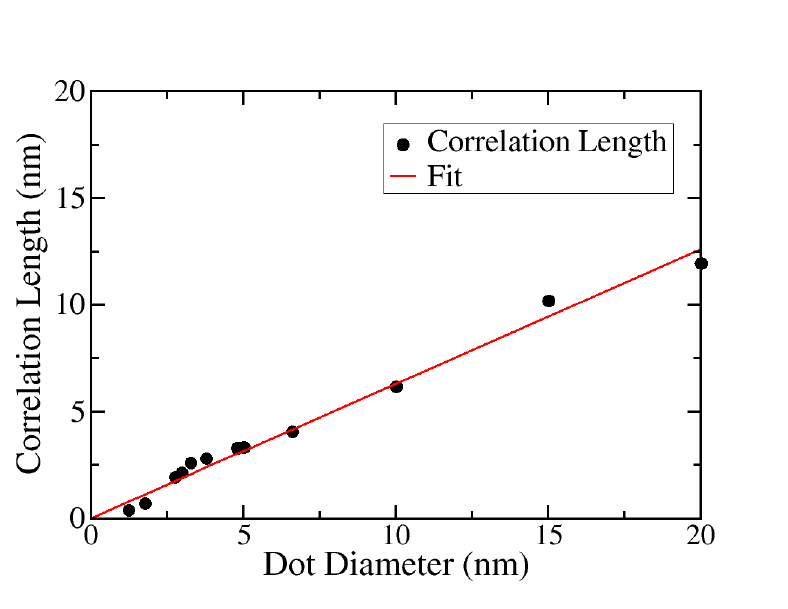}
\vspace{15 pt}
\end{wrapfigure}      
\textcolor{BlueViolet}{\textbf{ABSTRACT:}} 
The electron-hole correlation length serves as an intrinsic length
scale for analyzing excitonic interactions in semiconductor nanoparticles. 
In this work, the derivation of electron-hole correlation 
length using the two-particle reduced density is presented.
The correlation length was obtained by first calculating 
the electron-hole cumulant from the pair density,
and then transforming the cumulant into intracular coordinates, and finally then 
imposing exact sum-rule conditions on the radial integral of the cumulant.
The excitonic wave function for the calculation was obtained variationally using the electron-hole explicitly correlated Hartree-Fock method. 
 As a consequence, both the 
pair density and the cumulant were  explicit functions of 
the electron-hole separation distance. The use of explicitly correlated wave function 
and the  integral sum-rule condition are the two key features of this derivation. 
The method was applied to a series of CdSe quantum dots with diameters
1-20 nm and the effect of dot size on the correlation length was analyzed. 
}
}
\noindent\textcolor{BlueViolet}{\rule{14.6cm}{1.2pt}} 
\end{abstract}


\maketitle

\textcolor{BlueViolet}{\section{Introduction}}
Electron-hole excitations in semiconductor quantum dots 
are influenced by  their size, shape and chemical composition. Controlling the generation and the dissociation of electron-hole (eh) 
pairs have important technological applications in the field of
light-harvesting materials\cite{Zhu2011,Wu2013,Gratzel2005,Kamat2008}, 
photovoltaics\cite{Li2014,Golobostanfard2014,Mcdonald2005,Chetia2014}, solid-state lighting\cite{Anikeeva2007,Jang2010,Sohn2014,Shea-Rohwer2013} and
lasing\cite{Klimov2000,Foucher2014,Wu2014,Wang2014}. In order to control the generation and dissociation of the eh-pair, it is important to understand the underlying interaction between the quasiparticles.  
Theoretical treatment of electron-hole interaction in quantum dots
is challenging because of the computational bottleneck associated with 
quantum mechanical treatment of many-electron systems. In principle, a 
simplified description of the electron-hole pair can be achieved by ignoring the 
eh interaction and treating them as independent quasiparticles. 
Although this approach can dramatically reduce the computational cost,
such simplification can lead to qualitatively wrong results. 
For example, optical spectra calculation using independent quasiparticle
approach often shows significant deviation from the experimental results. One of the main limitations
of the independent quasiparticle method is its inability in describing bound 
excitonic states. Multiexcitonic interaction, exciton and biexciton binding energies,
radiative and Auger recombination are some of the properties whose calculations 
depend on the accurate treatment of electron-hole correlation.
Theoretical investigation of
electron-hole correlation has been performed using various
methods such as time-dependent density functional theory (TDDFT)\cite{doi:10.1021/jp209756z, Castro2012,Ullrich2000,C2NR11398H,doi:10.1021/nn2038884, doi:10.1021/jp206622e,Puerto2006,Troparevsky2003}, 
perturbation theory\cite{doi:10.1021/ct300946j}, 
 GW combined with 
Bethe-Salpeter equation\cite{Janis2012,Pal2010,Pal2011,Perebeinos2005,Puschnig2002,Rohlfing1998,raey,Ping2012,Rohlfing2000},
 configuration interaction\cite{Brasken2002,Corni2003,Corni2003a,Hu1990,He2005,Franceschetti2001,doi:10.1021/nl071219f,Rabani1999,Farmanzadeh2013}, quantum Monte Carlo\cite{Hu1990,Shumway2006,Shumway2005,Zhu1996},
path-integral Monte Carlo,\cite{Harowitz2005,Harowitz2005a}
explicitly correlated Hartree-Fock method,\cite{Elward2013,Elward2012,Elward2012a,Elward2012b,Blanton2013}
and electron-hole density functional theory.\cite{Sander1973}
\par
In this work, we are interested in the calculation of electron-hole correlation length (eh-CL) in CdSe quantum dots. 
Our goal is to provide a statistical definition of the electron-hole correlation length.
The concept of correlation length has been widely used in many 
fields, including statistical mechanics\cite{Brush1967,PhysRevB.65.134411,PhysRevLett.71.3198,PhysRevE.80.046708} and polymer science.\cite{Domb1972,Gujrati1998,Spouge1988,Vilenkin1978,PhysRevB.65.134411,PhysRevLett.71.3198,PhysRevE.80.046708} 
One of the important features of the eh-CL is that is it provides
an intrinsic length scale for describing the electron-hole interaction. Because of this,
it can play an important role in describing excitonic effects in quantum dots and 
other nanomaterials such as carbon nanotubes.\cite{Luer2009,Koch2003,Corfdir2013} 
The eh-CL can also be used for construction of electron-hole correlation functional 
for multicomponent density functional theory.\cite{Sander1973}
For example, Salahub and co-workers have developed a series of exchange-correlation 
functions that are based on electron-electron correlation length\cite{Proynov1994,Proynov1994a,Proynov1994b,Proynov1995} 
and a similar strategy 
can be used for construction of electron-hole correlation functionals using eh-CL. 
The eh-CL can also aid in the development of explicitly correlated wave functions (such as Jastrow and Gaussian-type geminal functions) 
which depend directly on  the electron-hole 
separation distance.\cite{Zhu1996,Elward2013,Elward2012,Elward2012a,Elward2012b,Blanton2013}
\par
We have used the 2-particle electron-hole density matrix for the definition and calculation of the eh-CL. 
Two-particle reduced density matrix (2-RDM) has been used extensively for investigation of electron-electron correlation\cite{Mazziotti2006,Mazziotti200721,Mazziotti2012244,Rohr2010,Rohr2011,Rajam2010,Elliott2011}
and electronic excitation\cite{Chatterjee2012} in many-electron systems. For the present system, the 2-RDM 
is the appropriate mathematical quantity that contains all the necessary information about electron-hole correlation. 
Specifically, the cumulant associated with the electron-hole 2-RDM is the component of the 2-RDM that cannot be expressed
as a product of 1-particle electron and hole densities. In principle, the 2-RDM can be obtained directly without the need for 
 an underlying wave function as long as the $N$-representability of 2-RDM can be satisfied. However, in the present work, 
we have obtained the 2-RDM from an explicitly correlated electron-hole wave function. 
The remainder of the article is organized as follows.
The derivation of eh-CL from the electron-hole cumulant is presented in \autoref{sec:cumulant},  
transformation to intracular and extracular coordinates is described in \autoref{sec:intra_coord}, and
details of the  explicitly correlated electron-hole wave function are presented in \autoref{sec:xchf} and \autoref{sec:comp}. 
The method was applied to a series of CdSe quantum dots and the results are presented in \autoref{sec:sec_results}.  
\textcolor{BlueViolet}{\section{Theory}\label{sec:sec_theory}}
\subsection{Electron-hole cumulant}
\label{sec:cumulant}
The interaction between the quasiparticles in the quantum dot is 
described the electron-hole Hamiltonian\cite{Zhu1996,Hu1990,Burovski2001, Wimmer2006,Woggon2013,PSSB:PSSB775,Corni2003,Corni2003a,Corni2003b,PSSB:PSSB200642169,Vanska2010,C2PC90004A,Blanton2013,Elward2013,Elward2012,Elward2012b} 
which has the following general expression
\begin{align}
\label{eq:heh}
	H &= 
	\sum_{ij} \langle i \vert 
	\frac{-\hbar^2}{2m_\mathrm{e}}\nabla^2_\mathrm{e} + v_\mathrm{ext}^\mathrm{e}
	\vert j\rangle e_i^\dagger e_j \\ \notag
    &+ \sum_{ij} \langle i \vert 
    \frac{-\hbar^2}{2m_\mathrm{h}}\nabla^2_\mathrm{h}+ v_\mathrm{ext}^\mathrm{h}
    \vert j\rangle     h^\dagger_i h_j \\ \notag
    &+ \sum_{iji'j'}    \langle iji'j' \vert
    \epsilon^{-1} r_\mathrm{eh}^{-1}
    \vert iji'j' \rangle
    e^\dagger_i e_j h_{i'}^\dagger h_{j'} \\ \notag
    &+\sum_{ijkl} w_{ijkl}^\mathrm{ee} e^\dagger_i e_j^\dagger e_l e_k 
    +\sum_{ijkl} w_{ijkl}^\mathrm{hh} h^\dagger_i h_j^\dagger h_l h_k.
\end{align}
We define the electron-hole wave function for a
multiexcitonic system consisting of $N_\mathrm{e}$
and $N_\mathrm{h}$ number of electrons and holes, respectively 
by 
$\Psi_\mathrm{eh}(
\mathbf{x}_1^\mathrm{e},\dots,\mathbf{x}_{N_\mathrm{e}}^\mathrm{e},
\mathbf{x}_1^\mathrm{h},\dots,\mathbf{x}_{N_\mathrm{h}}^\mathrm{h})$,
where $\mathbf{x}$ is a compact notation for both the spatial and spin coordinate
of the particles. 
The spin-integrated 2-particle reduced density can be obtained from the electron-hole
wave function by integration over the $N_\mathrm{e}-1$ and $N_\mathrm{h}-1$ coordinates as shown in
the following equation
\begin{align}
\label{eq:rho2}
	\rho_{\mathrm{eh}}(\mathbf{r}^{\mathrm{e}},\mathbf{r}^{\mathrm{h}})	
	&=
	\frac{N_\mathrm{e} N_\mathrm{h}}{\langle \Psi_\mathrm{eh} \vert \Psi_\mathrm{eh} \rangle}
	\int ds_1^\mathrm{e} ds_1^\mathrm{h} 
	d\mathbf{x}_2^\mathrm{e},\dots,\mathbf{x}_{N_\mathrm{e}}^\mathrm{e}
	\mathbf{x}_2^\mathrm{h},\dots,\mathbf{x}_{N_\mathrm{h}}^\mathrm{h}
	\Psi_\mathrm{eh}^\ast \Psi_\mathrm{eh}
\end{align}
where, integration over the spin coordinate $s_1$ is performed for both electron and hole. 
The single-particle density is obtained from the 2-particle density using the sum-rule condition\cite{parr1989density}
\begin{align}
\label{eq:rho1}
	\rho_{\mathrm{e}}(\mathbf{r}^{\mathrm{e}})	
	&= \frac{1}{N_\mathrm{h}}
	\int d\mathbf{r}^\mathrm{h} 
	\rho_{\mathrm{eh}}(\mathbf{r}^{\mathrm{e}},\mathbf{r}^{\mathrm{h}}), \\
	\rho_{\mathrm{h}}(\mathbf{r}^{\mathrm{h}})	
	&= \frac{1}{N_\mathrm{e}}
	\int d\mathbf{r}^\mathrm{e} 
	\rho_{\mathrm{eh}}(\mathbf{r}^{\mathrm{e}},\mathbf{r}^{\mathrm{h}}) .
\end{align} 
We define the electron-hole cumulant as the difference between the 2-particle density 
and the product of the 1-particle electron and hole densities as shown in the
following equation 
\begin{align}
	\label{eq:cumulant}
	q(\mathbf{r}^{\mathrm{e}},\mathbf{r}^{\mathrm{h}})	
	&=	
	\rho_{\mathrm{eh}}(\mathbf{r}^{\mathrm{e}},\mathbf{r}^{\mathrm{h}})	
	- \rho_{\mathrm{e}}(\mathbf{r}^{\mathrm{e}})
	\rho_{\mathrm{h}}(\mathbf{r}^{\mathrm{h}})	.
\end{align}
This definition is analogous to the definition used 
by Mazziotti et al.\cite{Juhasz2006}
in electronic structure theory. By construction,
the cumulant contains information about correlation between the 
two particles.  
Consequently, the Coulomb contribution of the electron-hole 
correlation energy can be directly expressed in terms of the 
electron-hole cumulant and is given by the following expression 
\begin{align}
	\label{eq:cumulantJeh}
	\langle \Psi_\mathrm{eh} \vert V_\mathrm{eh} \vert \Psi_\mathrm{eh} \rangle
	&=
	\langle \rho_\mathrm{eh} \epsilon^{-1} r_\mathrm{eh}^{-1} \rangle\\ 
	&=
	J_\mathrm{eh} + \langle q(\mathbf{r}^\mathrm{e},\mathbf{r}^\mathrm{h}) \epsilon^{-1} r_\mathrm{eh}^{-1} \rangle,\nonumber
\end{align}
where $\epsilon$ is the dielectric constant and  $J_\mathrm{eh}$ is the classical Coulomb electron-hole energy
\begin{align}
	\label{eq:coulomb}
	J_\mathrm{eh} 
	&= \langle \rho_\mathrm{e} \rho_\mathrm{h} \epsilon^{-1} r_\mathrm{eh}^{-1} \rangle.
\end{align}
The cumulant has an important property that its integration over all space should be 
zero due to the density sum-rule conditions\cite{parr1989density}
\begin{align}
\label{eq:qzero}
	\int d\mathbf{r}^\mathrm{e} d\mathbf{r}^\mathrm{h} q(\mathbf{r}^{\mathrm{e}},\mathbf{r}^{\mathrm{h}})
	&= 0 .
\end{align} 
We use this relationship for the definition of the electron-hole correlation 
length. 
\par
\subsection{Intracular and extracular coordinates}
\label{sec:intra_coord}
Beginning with Coleman's initial definition of the intracule and extacule matrices in terms 
of the center of mass (extracule) and relative motion (intracule) coordinates,\cite{Coleman1967}
the concept of the intracule and extracule in the regime of 
electronic systems has been previously explored in earlier studies.\cite{Coleman1967,Ugalde199151,Koga19983424,Gill2000303,Gill2003241,Besley20032033,Gill200615}
The intracular and extracular coordinates for 
the eh-system are defined by 
\begin{align}
	\label{eq:intracular}
	\mathbf{r}_\mathrm{eh} &= \mathbf{r}^\mathrm{e}-\mathbf{r}^\mathrm{h} \\
	\label{eq:extacular}
	\mathbf{R} &= \frac{1}{2} \left( \mathbf{r}^\mathrm{e}+\mathbf{r}^\mathrm{h} \right).
\end{align}
 The integral of the cumulant is expressed in terms of these coordinates
\begin{align}
 \int d\mathbf{r}^\mathrm{e} d\mathbf{r}^\mathrm{h} q(\mathbf{r}^{\mathrm{e}},\mathbf{r} ^{\mathrm{h}})
 &=
 \int d\mathbf{r}_\mathrm{eh}  \int d\mathbf{R} q(\mathbf{r}_\mathrm{eh},\mathbf{R}) \\
 &=
 \int_0^\infty d{r}_\mathrm{eh} r^2_\mathrm{eh} 
 \int d\Omega \sin\theta  \int d\mathbf{R} q(\mathbf{r}_\mathrm{eh},\mathbf{R}) \\
 &=
 \label{eq:radialcumulant}
 \int_0^\infty d{r}_\mathrm{eh} r^2_\mathrm{eh}  q_\mathrm{r}(r_\mathrm{eh})
\end{align}
In the above expression, the integral over the intracular coordinate $\mathbf{r}_\mathrm{eh}$ is transformed into spherical polar coordinates. The function $q_r$ is the spherically averaged radial cumulant and  the integral of the radial cumulant over a finite limit is used to define the following function $I(d)$ 
\begin{align}
 I(d) = \int_0^d dr_\mathrm{eh} r^2_\mathrm{eh} q_r(r_\mathrm{eh}).
 \label{eq:iofd}
\end{align}
The zero-integral property of $q$ (defined in Eq. \eqref{eq:qzero})  ensures that this integral goes to zero at large $d$
\begin{align}
	\lim_{d \rightarrow \infty} I(d) = 0 .
\end{align}
Here, we use $I(d)$ to define the electron-hole correlation length. 
Specifically, the electron-hole correlation length ($r_\mathrm{c}$) is defined as the 
value of $d$ at which the value of $I(d)$ is zero
\begin{align}
\label{eq:Ic}
 \vert I(r_\mathrm{c}) \vert &= 0 \quad \quad  r_\mathrm{c} << \infty .
\end{align}
The description of the 
electron-hole wave function used for the calculation of the radial cumulant is 
presented in the following section. 
\par
\subsection{Explicitly correlated electron-hole wave function}
\label{sec:xchf}
We have used the electron-hole explicitly correlated Hartree-Fock method (eh-XCHF)
for obtaining the electron-hole wave function. This method has been used in 
earlier work for the computation of exciton binding energies and electron-hole 
recombination probabilities in quantum dots.\cite{Elward2012,Elward2012a,Elward2012b,Elward2013,Blanton2013} 
A brief summary of the eh-XCHF method is presented here and the implementation 
details of this method can be found in work by Elward and 
co-workers.\cite{Elward2012,Elward2012a,Elward2012b} 
The ansatz of the  eh-XCHF wave function consists of multiplying the 
mean-field electron-hole reference wave functions with an explicitly correlated 
function $G$ as shown in the following equation
\begin{align}
\label{eq:xchfansatz}
\Psi_{\mathrm{eh-XCHF}}= G \Phi_{\mathrm{e}}\Phi_{\mathrm{h}} ,
\end{align} 
where $G$ is the geminal operator
\begin{align}
	\label{eq:gemdef}
	G &= 
	\sum_{i=1}^{N_{\mathrm{{e}}}} 
	\sum_{j=1}^{N_{\mathrm{{h}}}} g(r_{ij})   , \\
	g(r_\mathrm{eh}) &= \sum_{k=1}^{N_{\mathrm{{g}}}}
	b_k \exp(-\gamma_k r_{\mathrm{eh}}^2).
\end{align}
The eh-XCHF method is a variational method in which the correlation function $G$ and the
reference wave function are obtained by minimizing the total energy
\begin{align}
	\label{eq:xchfmin}
	E_\mathrm{eh-XCHF}
	&=
	\min_{G,\Phi_\mathrm{e},\Phi_\mathrm{h}}
	\frac{\langle G \Phi_0\vert H \vert G \Phi_0\rangle}{\langle G \Phi_0\vert G \Phi_0\rangle},
\end{align} 
where $\Phi_0=\Phi_\mathrm{e}\Phi_\mathrm{h}$.
To perform the above minimization, it is more efficient to work with the following
congruent-transformed operators
\begin{align}
\label{eq:tilde}
 \tilde{H} &= G^\dagger H G, \\
 \tilde{1} &= G^\dagger G. 
\end{align}
This transformation is particularly important for the calculation of the 
2-particle reduced density matrix in the present work. 
The set of parameters $\{b_k,\gamma_k\}$ in $G$ were obtained by non-linear optimization, and for a given set of these parameters, the minimization over the reference wave function was performed 
by determining the self-consistent solution of the coupled Fock equations
\begin{align}
\label{eq:fock}
	\tilde{\mathbf{F}}_\mathrm{e} \mathbf{C}_\mathrm{e} &= \tilde{\mathbf{S}}_\mathrm{e} \mathbf{C}_\mathrm{e} \lambda_\mathrm{e}, \\
	\tilde{\mathbf{F}}_\mathrm{h} \mathbf{C}_\mathrm{h} &= \tilde{\mathbf{S}}_\mathrm{h} \mathbf{C}_\mathrm{h} \lambda_\mathrm{h}.
\end{align}
The tilde in the above expressions represent that the Fock and the overlap matrices incorporate the transformed operators defined in Eq. \eqref{eq:tilde}. 
\par
The transformed operator $\tilde{1}$ can be written as a sum of operators as shown below
\begin{align}
	\tilde{1} 
	&= 
		G^\dagger G \\
	&= 
		\sum_{ii'} g(i,i') \sum_{jj'} g(j,j') 
	   \quad (i,j=1,\dots,N_\mathrm{e} ; i',j'=1,\dots,N_\mathrm{h}) \\
	&=
		\sum_{ii'} g(i,i')g(i,i') 
    +	\sum_{i \neq j,i'} g(i,i')g(j,i') \\ \notag
    &+  \sum_{i' \neq j',i} g(i,i')g(i,j')
    +	\sum_{i \neq j,i' \neq j'} g(i,i')g(j,j').
\end{align}
The above expression can be written in a compact notation as a sum of 
2, 3, and 4-particle operators
\begin{align}
\label{eq:omega}
	G^\dagger G
	&=
	\Omega_{11} + \Omega_{21} + \Omega_{12} + \Omega_{22} .
\end{align} 
The 2-particle density for the eh-XCHF wave function can be expressed in terms of these
operators as shown below
\begin{align}
	\rho_{\mathrm{eh}}(\mathbf{r}^{\mathrm{e}},\mathbf{r}^{\mathrm{h}})	
	&=
	\frac{N_\mathrm{e} N_\mathrm{h}}{\langle \Psi_\mathrm{eh-XCHF} \vert \Psi_\mathrm{eh-XCHF} \rangle} \\ \notag
	& \times
	\langle
	\Psi_\mathrm{eh-XCHF}^\ast \Psi_\mathrm{eh-XCHF} 
	\rangle_{s_1,s_1',2,2',\dots,N_\mathrm{e},N_\mathrm{h}},
\end{align}
where the subscript in the above expression is a compact notation for 
integration over the remaining coordinates described in Eq. \eqref{eq:rho2}. 
Substituting the expression from Eq. \eqref{eq:omega}, we get the following 
expression
\begin{align}
\label{eq:rho_omega}
	\rho_{\mathrm{eh}}(\mathbf{r}^{\mathrm{e}},\mathbf{r}^{\mathrm{h}})	
	&=
	\frac{N_\mathrm{e} N_\mathrm{h}}{\langle \Phi_0 \vert \tilde{1} \vert \Phi_0 \rangle} \\ \notag
	& \times
	\langle
	\Phi_0^\ast (\Omega_{11}+\Omega_{12}+\Omega_{21}+\Omega_{22}) \Phi_0 
	\rangle_{s_1,s_1',2,2',\dots,N_\mathrm{e},N_\mathrm{h}}.	
\end{align}
For a multiexcitonic system all 2, 3, and 4-particle operators should be used for 
the computation of the 2-particle density. In a related work on many-electron system, we have shown that it is possible to avoid integration over higher-order operators by using
diagrammatic summation technique and a similar strategy can be used for multiexcitonic systems as well.\cite{Bayne2014}
\par
\subsection{Relation to uncorrelated transition density matrices}
\label{sec:dmat}
\begin{figure}[h!]
  \begin{center}
    \includegraphics{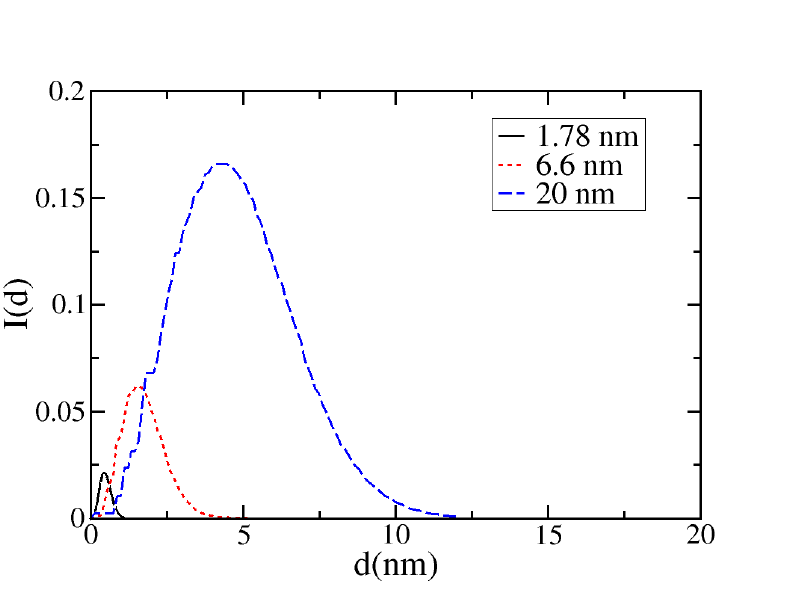}
    \caption{The value of $I(d)$ as $d$, the upper limit in Eq. \eqref{eq:iofd}, is varied for the 1.78 nm, 6.6 nm, and 20 nm diameter CdSe quantum dots.}
    \label{fig:intcorrfuncs}
   \end{center}
\end{figure}

\begin{figure}[h!]
  \begin{center}
        \includegraphics{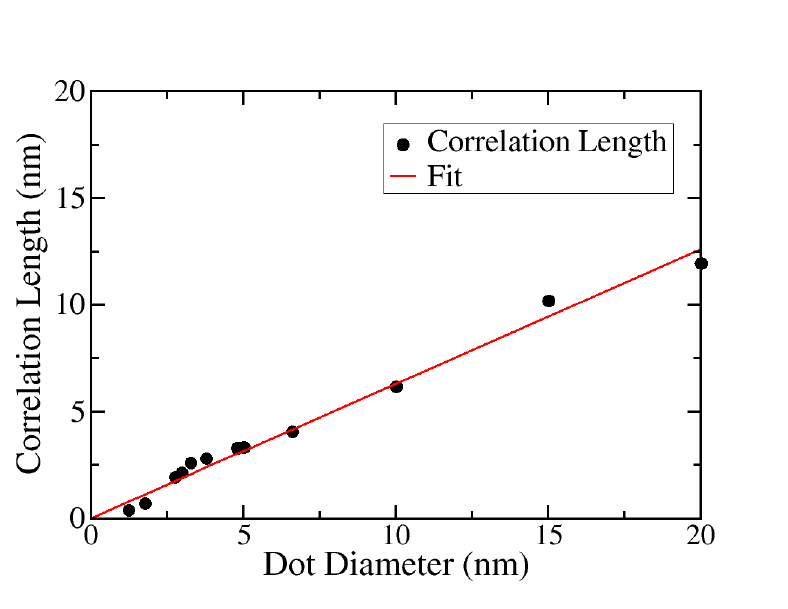}
    \caption{The correlation length as a function of the dot diameter $D_{\mathrm{dot}}$. The linear fit is $r_{\mathrm{C}}=0.6303D_{\mathrm{dot}}$, with a mean absolute error of 0.323 nm.}
    \label{fig:corrlengths}
  \end{center}
\end{figure}

One of the important features of the correlation function is that it allows for a
compact representation of the 2-particle density matrix in the position representation. 
The relationship can be readily seen by expanding the eh-XCHF wave function in the Slater
determinant basis
\begin{align}
	\label{eq:Gphi0trans}
	G \Phi_0  
	= 
	\sum_{ii'}^\infty
	\underbrace{• 
	\langle \Phi_i^\mathrm{e} \Phi_{i'}^\mathrm{h} \vert G \vert \Phi_0^\mathrm{e} 
	\Phi_0^\mathrm{h} \rangle }_{c_{ii'}}
	 \Phi_i^\mathrm{e} \Phi_{i'}^\mathrm{h} 
	=
	\sum_{ii'}^\infty c_{ii'}  \Phi_i^\mathrm{e} \Phi_{i'}^\mathrm{h}. 
\end{align}
Substituting Eq.~\eqref{eq:Gphi0trans} in the expression of $\rho_\mathrm{eh}$ gives
\begin{align}
\label{eq:rho_ci}
	\rho_{\mathrm{eh}}(\mathbf{r}^{\mathrm{e}},\mathbf{r}^{\mathrm{h}})	
	&=
	\frac{N_\mathrm{e} N_\mathrm{h}}{\langle \Phi_0 \vert \tilde{1} \vert \Phi_0 \rangle} \\ \notag
	& \times
	\langle
	\sum_{ij}^\infty \sum_{i'j'}^\infty c_{ii'}^\ast c_{jj'}
	\Phi_i^{\mathrm{e}\ast} \Phi_{i'}^{\mathrm{h}\ast}
	\Phi_j^{\mathrm{e}} \Phi_{j'}^{\mathrm{h}}
	\rangle_{s_1,s_1',2,2',\dots,N_\mathrm{e},N_\mathrm{h}}	 \\
	&=
	\frac{N_\mathrm{e} N_\mathrm{h}}{\langle \Phi_0 \vert \tilde{1} \vert \Phi_0 \rangle} 
	\sum_{ij}^\infty \sum_{i'j'}^\infty c_{ii'}^\ast c_{jj'}
	d_{ij}^\mathrm{e} d_{i'j'}^\mathrm{h},
\end{align}
where the transition density matrix $d_{ij}$ is defined as
\begin{align}
	\label{eq:transdensmat}
	d_{ij}^\mathrm{e} (\mathbf{r}^\mathrm{e})
	&=
	\langle
	\Phi_i^{\mathrm{e}\ast} \Phi_{j}^{\mathrm{e}}
	\rangle_{s_1,2,\dots,N_\mathrm{e}}.
\end{align}
It is seen from Eq. \eqref{eq:rho_ci} that the 2-particle density obtained from the 
eh-XCHF wave function is equivalent to the infinite-order expansion in terms of the 
transition density matrices. 
\textcolor{BlueViolet}{\section{Computational details}\label{sec:comp}}
The method described in \autoref{sec:sec_theory} was used for calculating electron-hole correlation
length in CdSe quantum dots in the range of 1-20 nm in diameter. 
We are interested in the effect of dot size on the electron-hole correlation length 
for a  single electron-hole pair in CdSe quantum dots. For a single electron-hole pair, 
the higher-order operators in Eq. \eqref{eq:rho_omega}  rigorously vanish from the expression.  
This provides considerable simplification in the calculation of the 2-particle density. 
Because of the dot size, 
application of either DFT or atom-centered pseudopotential approach is computationally 
prohibitive. To make the computation tractable, we have used a parabolic confining potential 
in the electron-hole Hamiltonian described in Eq. \eqref{eq:heh}. Parabolic confinement potential 
in quantum dots has been used extensively for various
 properties such as 
 total exciton energy\cite{El-Said1994,Que199211036},
 exciton dissociation\cite{Nenashev2011}, 
 exciton binding energy\cite{Poszwa2010,Elward2012,Elward2012b,Elward2013} eh-recombination
probability\cite{Elward2012,Elward2012a,Elward2013}, effect of magnetic\cite{Jaziri1994,Halonen19925980,Song20011253021,Taut2009,Kolovsky2014,Trojnar2011} and electric fields\cite{Jaziri1994,Xie2009,He2010,Blanton2013,Fernandez-Menchero2013},
exciton-polariton condensate\cite{Fernandez2013},
linear optical properties\cite{Rey2005,Kim2007},
optical rectification\cite{Rezaei2011},
non-linear rectification\cite{Xie2009}, dynamics\cite{Liu2011}, eh-correlation energy\cite{Blundell2010,Zhao2011}, resonant tunneling\cite{Teichmann20133571}, collective modes\cite{Morales2008}, and thermodynamic 
properties\cite{Nammas2011}.
The external potential for the electron and hole quasiparticle was defined as
\begin{align}
	v_\alpha^\mathrm{ext} = \frac{1}{2} k_\alpha \vert \mathbf{r}_\alpha\vert^2 \quad \alpha = \mathrm{e,h}	
\end{align}
where $k_\alpha$ is the force constant which determines the strength of the 
confinement potential. We have used a particle-number based search procedure for 
determination of the force constant $k_\alpha$.  The central idea of this approach is 
to find the value of $k_\alpha$ such that the computed 1-particle electron and hole densities 
are confined within the volume of the quantum dot. This is obtained by performing the 
following minimization
\begin{align}
	\min_{k_\alpha^\mathrm{min}}
	\left(
	N_\alpha - \int_0^{\frac{D_\mathrm{dot}}{2}} dr r^2 \int d\Omega  \rho_\alpha(\mathbf{r})
	\right)^2,
\end{align}
where $D_\mathrm{dot}$ is the diameter of the quantum dot and $\Omega$ is the angular coordinate.
The values of the force constants used for each dot is listed in  \autoref{tab:forceconstants}.
\begin{table}
	\begin{center}
	\caption{\textbf{Force constants for CdSe quantum dots.}}
	\label{tab:forceconstants}
	\begin{ruledtabular}
	\begin{tabular}{d M M}
	\multicolumn{1}{c}{Dot diameter (nm)} & \multicolumn{1}{c}{$k_\mathrm{e}$ (atomic units)}& \multicolumn{1}{c}{$k_\mathrm{h}$ (atomic units)}\\
	\hline
	1.24 & 2.66\times 10^{-2} & 9.10\times 10^{-3}\\
	1.79 & 6.22\times 10^{-3} & 2.13\times 10^{-3}\\
	2.76 & 1.10\times 10^{-3} & 3.76\times 10^{-4}\\
	2.98 & 8.10\times 10^{-4} & 2.77\times 10^{-4}\\
	3.28 & 5.52\times 10^{-4} & 1.89\times 10^{-4}\\
	3.79 & 3.09\times 10^{-4} & 1.06\times 10^{-4}\\
	4.80 & 1.20\times 10^{-4} & 4.12\times 10^{-5}\\
	5.00 & 1.02\times 10^{-4} & 3.51\times 10^{-5}\\
	6.60 & 3.40\times 10^{-5} & 1.16\times 10^{-5}\\
	10.00 & 6.41\times 10^{-6} & 2.19\times 10^{-6}\\
	15.00 & 1.26\times 10^{-6} & 4.33\times 10^{-7}\\
	20.00 & 4.01\times 10^{-7} & 1.37\times 10^{-7}\\
	\end{tabular}
	\end{ruledtabular}
	\end{center}
\end{table}
The kinetic energy operator was computed using the electron and hole effective masses 
of $0.13$ and $0.38$ atomic units, respectively.\cite{Wimmer2006}
The interaction between the electron and hole was 
described by screened Coulomb potential. We have used the size and distance dependent dielectric function $\epsilon(\mathbf{r},R_\mathrm{dot})$, which was developed by Wang and  Zunger for CdSe.\cite{Wang1996}
The electron and hole molecular orbitals in $\Phi_0$ were represented using 
a linear combination of Gaussian type orbitals (GTOs) and the expansion 
coefficients were obtained by the solving the coupled Fock equations 
shown in Eq. \eqref{eq:fock}. The basis used was a single S Cartesian GTO was used and the 
exponents of the basis functions are listed in \autoref{tab:basis}.
\begin{table}[ht]
  \begin{center}
   \caption{\textbf{Exponent used in GTO basis $e^{-\alpha r^2}$.}}
   \label{tab:basis}
   \begin{ruledtabular}
    \begin{tabular}{ d M }  
      \multicolumn{1}{c}{Dot diameter (nm)}   &  \multicolumn{1}{c}{$\alpha$ (atomic units)} \\
     \hline 
        1.24  &  2.94\times 10^{-2}  \\
        1.78   &    1.42\times 10^{-2}  \\
        2.76    &   5.98\times 10^{-2}  \\
        2.98    &   5.13\times 10^{-2}  \\
        3.28    &   4.24\times 10^{-3}  \\
        3.79    &   3.17\times 10^{-3}  \\
        4.80    &   1.98\times 10^{-3}  \\
        5.00    &   1.83\times 10^{-3}  \\
        6.60    &   1.05\times 10^{-3}  \\
        10.00    &   4.57\times 10^{-4}  \\
        15.00    &   2.03\times 10^{-4}  \\
        20.00    &   1.14\times 10^{-4}        
    \end{tabular}
    \end{ruledtabular}
  \end{center}  
\end{table}   
The use of GTOs is especially convenient because the 
integrals involving the GTOs and the Gaussian correlation function, $G$, are known 
analytically.\cite{Boys1950,Singer1960,Persson1996,Persson1997} 
For a given value of $\mathbf{r}^\mathrm{e}$, the 1-particle density 
$\rho$ was calculated analytically. The integration over the intracular coordinate in Eq. \eqref{eq:iofd}
was performed numerically. 
The correlation function, $G$, was expanded as a linear combination of 
six Guassian-type geminal functions\cite{Elward2012,Elward2012b,Elward2013} and the set of $\{b_k,\gamma_k\}$
parameters were optimized for each dot size. The first 
set of geminal parameters was set to $b_1=1$ and $\gamma_1=0$ for all CdSe 
dot diameters. For each dot diameter, five sets of geminal parameters were 
determined 
sequentially by minimizing  the energy. 
The values of the geminal parameters are found in \autoref{tab:geminals}. 
{\squeezetable
\begin{table*}[ht]
  \begin{center}
   \caption{\textbf{Value of the geminal parameters for CdSe quantum dots.}}
   \label{tab:geminals}
   \begin{ruledtabular}
    \begin{tabular}{ d  M M  M  M  M  M  M  M  M  M}  
      \multicolumn{1}{c}{Dot Diameter (nm)}   &  \multicolumn{1}{c}{$b_2$} & \multicolumn{1}{c}{$\gamma_2$} & \multicolumn{1}{c}{$b_3$} & \multicolumn{1}{c}{$\gamma_3$} & \multicolumn{1}{c}{$b_4$} & \multicolumn{1}{c}{$\gamma_4$} & \multicolumn{1}{c}{$b_5$} & \multicolumn{1}{c}{$\gamma_5$} & \multicolumn{1}{c}{$b_6$} & \multicolumn{1}{c}{$\gamma_6$} \\
     \hline 
        1.25  &  1.338\times 10^{-1}  & 2.134\times 10^{-2} & 1.559\times 10^{-1} & 1.150\times 10^{-3} & 3.630\times 10^{-2} & 1.112\times 10^{0} & 2.290\times 10^{-2} & 2.111\times 10^{-1} & 1.099\times 10^{-1} & 1.020\times 10^{-3} \\
        1.78   &  1.497\times 10^{-1} & 1.569\times 10^{-2} & 2.099\times 10^{-1} & 1.120\times 10^{-3} & 2.009\times 10^{-1} & 4.400\times 10^{-4} & 4.260\times 10^{-2} & 1.112\times  10^{0} & 3.820\times 10^{-2} & 1.444\times 10^{-1}  \\
        2.76    & 2.279\times 10^{-1} & 4.700\times 10^{-3}& 5.990\times 10^{-2}& 1.111\times 10^{-1}& 2.119\times 10^{-1}&4.400\times 10^{-4} & -1.910\times 10^{-2}& 3.140\times 10^{-3}& 1.540\times 10^{-2}&1.222\times 10^{0}   \\
        2.98    &2.449\times 10^{-1}  & 3.960\times 10^{-3}& 6.400\times 10^{-2}& 1.114\times 10^{-1}& 2.099\times 10^{-1}& 3.700\times 10^{-4}& 1.490\times 10^{-2}& 1.434\times 10^{0}& 1.580\times 10^{-2}& 1.012\times 10^{-1}  \\
        3.28    &2.589\times 10^{-1}  &3.580\times 10^{-3} &6.680\times 10^{-2} &1.112\times 10^{-1} &2.109\times 10^{-1} &4.300\times 10^{-4} &9.990\times 10^{-2} &1.100\times 10^{-4} &1.590\times 10^{-2} & 1.253\times 10^{0}  \\
        3.79    &2.799\times 10^{-1}  &3.000\times 10^{-3} &6.990\times 10^{-2} &1.111\times 10^{-1} &2.209\times 10^{-1} &3.700\times 10^{-4} &1.099\times 10^{-1} &1.200\times 10^{-4} &1.540\times 10^{-2} &1.432\times 10^{0}   \\
        4.80    &3.610\times 10^{-1}  &1.660\times 10^{-3} &8.790\times 10^{-2} &1.010\times 10^{-1} &6.180\times 10^{-2} &2.229\times 10^{-2} &2.099\times 10^{-1} &2.200\times 10^{-4} &9.990\times 10^{-2} &1.100\times 10^{-4}   \\
        5.00    &3.699\times 10^{-1}  &1.570\times 10^{-3} &8.990\times 10^{-2} &1.010\times 10^{-1} &6.680\times 10^{-2} &2.135\times 10^{-2} &2.099\times 10^{-1} &2.200\times 10^{-4} &9.990\times 10^{-2} &1.100\times 10^{-4}   \\
        6.60    &4.499\times 10^{-1}  &1.380\times 10^{-3} &6.099\times 10^{-1} &1.800\times 10^{-4} &5.280\times 10^{-2} &1.112\times 10^{0} &1.299\times 10^{-1} &1.011\times 10^{-1} &1.331\times 10^{-1} &1.335\times 10^{-2}   \\
        10.00  &5.899\times 10^{-1}  &1.240\times 10^{-3} &9.999\times 10^{-1} &1.300\times 10^{-4} &6.400\times 10^{-2} &1.012\times 10^{0} &1.569\times 10^{-1} &1.023\times 10^{-1} &1.799\times 10^{-1} &1.211\times 10^{-2}   \\
        15.00   &6.999\times 10^{-1} &1.040\times 10^{-3} &1.110\times 10^{0} &1.300\times 10^{-4} &6.580\times 10^{-2} &1.102\times 10^{0} &1.589\times 10^{-1} &1.022\times 10^{-1} &1.999\times 10^{-1} &1.999\times 10^{-1}   \\
        20.00    &7.999\times 10^{-1} &1.030\times 10^{-3} &2.000\times 10^{0} &1.200\times 10^{ -4} &9.999\times 10^{-1} &4.000\times 10^{-5} &5.489\times 10^{-1} &1.227\times 10^{-2} &2.899\times 10^{-1} &1.240\times 10^{-3}   \\
        
    \end{tabular}
    \end{ruledtabular}
  \end{center}  
\end{table*}
} 
\textcolor{BlueViolet}{\section{Results}\label{sec:sec_results}}
The electron-hole correlation length was obtained by integration of the radial cumulant as 
described in Eq. \eqref{eq:iofd}. In \autoref{fig:intcorrfuncs}, the integral of the cumulant, $I(d)$, 
for three different dot sizes are presented.
As expected, the integral goes to zero at large distances (high $d$ values) and the 
distance at which the integral converges to zero is defined as the 
electron-hole correlation length $r_\mathrm{c}$. 
The calculated electron-hole correlation lengths are presented in  \autoref{tab:corrlengths}.
\begin{table}
  \begin{center}
  \caption{\textbf{Electron-hole correlation lengths and $r_\mathrm{node}$ for CdSe quantum dots.}}
  \label{tab:corrlengths}
  \begin{ruledtabular}
  \begin{tabular}{d d d}
	 \multicolumn{1}{c}{Dot Diameter } & \multicolumn{1}{c}{Correlation length } & \multicolumn{1}{c}{$r_\mathrm{node}$ }\\
	 \multicolumn{1}{c}{(nm)} & \multicolumn{1}{c}{ (nm)} & \multicolumn{1}{c}{ (nm)}\\
	 \hline
	1.24 & 0.381 & 0.283\\
	1.78 & 0.683 & 0.431\\
	2.76 & 1.905 & 0.595\\
	2.98 & 2.117 & 0.653\\
	3.28 & 2.572 & 0.732\\
	3.79 & 2.778 & 0.833\\
	4.80 & 3.293 & 1.082\\
	5.00 & 3.307 & 1.124\\
	6.60 & 4.047 & 1.653\\
	10.00 & 6.156 & 2.749\\
	15.00 & 10.164 & 3.257\\
	20.00 & 11.930 & 4.733\\
  \end{tabular}
  \end{ruledtabular}
  \end{center}
\end{table}
 We find that, in all cases, the correlation length increases with increasing dot diameter.
Another quantity that is important for investigating electron-hole correlation is the 
length scale associated with the first node of the radial cumulant. We define this 
quantity as $r_\mathrm{node}$ and the calculated values are presented in \autoref{tab:corrlengths}.
The maximum of the $I(d)$ in \autoref{fig:intcorrfuncs} corresponds to $r_\mathrm{node}$.
Because the interaction between the electron and hole is attractive, we expect an enhancement in the 
pair density as compared to mean-field density at small $r_\mathrm{eh}$ distances. This phenomenon is 
opposite to the correlation hole observed in electron-electron interaction, in which small $r_\mathrm{ee}$ shows a decrease in correlated electron-pair density as compared to uncorrelated electron density. The $r_\mathrm{node}$
can be interpreted as the effective radius of the sphere that encloses the region of enhanced probability density. As seen from \autoref{tab:corrlengths}, $r_\mathrm{c}$ and $r_\mathrm{node}$ are similar in magnitude for  small dot sizes, but these quantities differ significantly for larger dots. 
The correlation
length as a function of the dot diameter is plotted in  \autoref{fig:corrlengths}. The set of data showed good agreement with the linear fit, with a mean absolute error of 0.323 nm. A trend of increasing correlation length with increasing dot
diameter is observed. The correlation lengths show that correlation effects are 
important even at long electron-hole separations. 
The linear relationship between the dot diameter and the correlation length has an important application 
in the construction of compact explicitly correlated electron-hole wave function.
For example, the determination of the eh-XCHF wave function requires the optimization of the non-linear parameters
 $\{b_k,\gamma_k\}$ in $G$. By using the relationship between the dot diameter and correlation length, it is
possible to assign the non-linear parameters as some multiple of the correlation length. 
This approach avoids optimization of non-linear parameters and can result in significant reduction in the computational effort.  
\textcolor{BlueViolet}{\section{Conclusions}\label{sec:sec_conclusion}}
In conclusion, we have presented a method for calculating electron-hole correlation length in semiconductor
quantum dots. We have used the cumulant derived from the electron-hole 2-particle density as the central quantity for defining the correlation length. There are two key features of this method. First, the 
2-particle reduced density was obtained from an explicitly correlated electron-hole wave function. 
Consequently, the reduced density matrix and the corresponding cumulant were explicit functions of the 
electron-hole separation distance. Second, the calculation of the correlation length was
not based on the nodes of the cumulant but was derived from  the 
exact sum rule relationship satisfied by all $N$-representable cumulants. The developed method 
was applied to a series of CdSe quantum dots and a linear relationship between the dot size and 
correlation length was observed. 
The electron-hole correlation length provides a natural length scale for investigating electron-hole
correlation in nanoparticles. We envision that in future work, the electron-hole correlation 
length will be used in the construction of compact explicitly correlated wave functions and also for developing 
multi-component\cite{Sander1973} electron-hole density functionals. 
\textcolor{BlueViolet}{\section*{Acknowledgments}
\label{sec:Acknowlegements}}
We wish to thank ACS-PRF  grant 52659-DNI6 and Syracuse University for financial support.  
\newpage
%
\end{document}